\documentclass[preprint]{jfm}
\usepackage{graphicx}%
\usepackage{subfigure}
\usepackage{epstopdf, epsfig}
\usepackage{dcolumn}%
\usepackage{bm}%
\usepackage{amsmath}
\usepackage{mathtools}

\shorttitle{Guidelines for authors}
\shorttitle{Velocity Spectra and Model Spectrum in Non-Premixed Jet Flames}

\shortauthor{Shamooni et al.}

\title{Velocity Spectra and Model Spectrum in Non-Premixed Jet Flames}

\author{Ali Shamooni\aff{1,2}
  \corresp{\email{ali.shamooni@polimi.it}}, Alberto Cuoci\aff{1}, Tiziano Faravelli\aff{1} \and \\ Amsini Sadiki\aff{2}
 }
 
\affiliation{\aff{1}Creck Modelling Lab, Department of Chemistry, Materials, and Chemical Engineering, Piazza Leonardo da Vinci 32, 20133 Milano, Italy 
\aff{2}Institute of Energy and Power Plant Technology, Technische Universit\"{a}t Darmstadt, 64287 Darmstadt,
Germany}

\begin{document}

\maketitle

\begin{abstract}
In this contribution, velocity and its dissipation spectra in turbulent non-premixed jet flames are investigated by using two Direct Numerical Simulation (DNS) databases of temporally evolving jets with different bulk jet Reynolds numbers. In the DNS differential diffusion and flame dynamics effects have been accounted for and the flames experience high levels of extinction followed by re-ignition. 
%The objective is to check Kolmogorov's scaling laws (with Favre averaged quantities) and to see if the dissipation spectra follow the model spectrum of Pope, proposed for non-reactive flows. 
It turns out that the spectra extracted from different statistically homogeneous planes across the jets and selected times corresponding to various occurring flame regimes (extinction and re-ignition phases) collapse reasonably when normalized by Favre averaged turbulent quantities. 
Especially (1) in the inertial range, the $\kappa^{-5/3}$ power law is observed with the constant of proportionality of $C_{K}=2.3$. 
(2) In the dissipation range $\exp(\beta\kappa)$ scaling exists with $\beta=7.2$ (instead of $\beta=5.2$).
(3) The location of the peak of the normalized 1D dissipation spectra is in a lower normalized waver number ($\approx 0.08$) compared to the peak of non-reactive model spectrum ($\approx 0.11$).
(4) Finally, an adapted 3D dissipation model spectrum proposed which peaks at a normalized waver number $\approx 0.19$ (instead of $\approx 0.26$).
%It was observed that the spectra, extracted from different planes across the jets and selected times corresponding to different flame regimes (extinction and re-ignition phases) collapse reasonably when normalized by Favre averaged turbulent quantities. A different exponential drop-off compared to the well known $\beta\approx 5.2$ in non-reactive regimes was detected. The location of the peak of the 1D dissipation spectra is in a lower normalized waver number ($\approx 0.08$) compared to the 1D model dissipation spectrum of Pope ($\approx 0.11$). If the observed $\beta \approx 7.2$, Kolmogorov's constant $C_K \approx 2.3$, and a fitted $c_\eta=0.28$ are replaced in the model spectrum, the reactive DNS data follows the model. The final adapted 3D dissipation model spectrum peaks at $\approx 0.19$ instead of $\approx 0.26$.
\end{abstract} 

\section{Introduction}
The analysis of velocity and scalar spectra is of great importance in both  experiments and numerics. In experiments, one needs to know true resolutions required to measure scalar gradients. %(\citet{wang2007dissipation}). 
On the other hand, many numerical models rely on scaling laws for the spectra to evaluate model coefficients or to justify assumptions, see e.g., \citet{Pope2000turbulent}. For non-reactive constant density incompressible flows a universal form of the normalized turbulent kinetic energy (TKE) spectrum function exists. The universal TKE spectrum function is $E_{Normal}(\kappa \underline{\eta})=\left(\kappa \underline{\eta}  \right)^{-5/3} C_K$, with $\kappa$ the wavenumber magnitude, $\underline{\eta}$ the Reynolds averaged Kolmogorov length scale, and $C_K$ the Kolmogorov constant (see e.g., \citet{saddoughi1994local} and for a review \citet{Sreenivasan1995}). 
%or \citet{nelkin1994universality}). 
%although some small deviations from the universality found to exist. 
Many works have been carried out so far on passive scalar spectra (see e.g.,  \citet{Batchelor1959}, \citet{Kraichnan1968}, and also 
% \citet{Corrsin1951}, \citet{Batchelor1959}, \citet{Kraichnan1968}, and also 
%\citet{Gibson1963}  and the reviews in   \citet{warhaft2000passive,sreenivasan1996passive}). 
%and Chapter 3 of  \citet{Davidson2012}). 
 a review in   \citet{warhaft2000passive}). 
Beside different behaviours which are observed in shear-less or shear flows, the scalar spectrum depends on the Schmidt (${\mathrm{Sc}}$) and the Taylor scale Reynolds numbers (${Re_{\lambda}}$), see e.g., \citet{donzis2010batchelor}, \citet{attili2013fluctuations}, \citet{yeung2013spectrum} or \citet{sreenivasan1996passive}. In general, the passive scalar spectrum behaviour is more complicated than the TKE (or in general the velocity spectrum). The focus of the current study is on the velocity spectrum.   
%It depends on the This spectrum depends on the relative importance of the viscous dissipation (as characterized by the kinematic viscosity ν) compared to the scalar diffusivity (κ), Schmidt number ${\mathrm{Sc}}$, the type of flow and the ${Re_{\lambda}}$. 
%In shear-less flows (where large scale anisotropy does not exist) the Schmidt number (${\mathrm{Sc}}$) dependency exists. 
%In $\mathrm{Sc}\leqq1$ and large Taylor Reynolds numbers (${Re_{\lambda}}$), by the dimensional analysis, the scalar fluctuations spectra have an approximate universal form: $E(\kappa)=C_{OC}\underline{\chi}\underline{\varepsilon}^{-1/3}\kappa^{-5/3}$, where $C_{OC}$ is the Obukhov-Corsin contant, with $\underline{\chi}$ the mean scalar dissipation rate. If $\mathrm{Sc}\gg1 $, the $\kappa^{-1}$ power form exists \citet{Batchelor1959} (also see the DNS results for the support \citet{donzis2010batchelor}). In shear flows where large scale anisotropy exist, the form of the scalar spectrum also depends on the shear length scale 
% \citet{attili2013fluctuations} and  ${Re_{\lambda}}$ \citet{sreenivasan1996passive}. The passive scalar spectrum is more complicated than the TKE (in general the velocity spectrum).  The current work is focused on the latter.  
 
Knowing that chemical reactions mostly occur in small scales, the kinetic energy and its dissipation spectra in high wavenumber ranges are of great importance. For reactive flows, the focus in the literature has been mainly on scalars spectra. As an example, experiments have been carried out in Sandia National Laboratories focusing on temperature and mixture fraction spectra of non-premixed jet flames C, D, and E and DLR-A and DLR-B (\citet{wang2007dissipation}). Scalars energy and their dissipation spectra were also reported in ICE engines (\citet{petersen2011high}), $\mathrm{CH_4/H_2/N_2}$ non-premixed jet flames 
%(\citet{frank2008high,wang2007quantification}) and DME/air partially premixed jet 
(\citet{wang2007quantification}) and DME/air partially premixed jet flames (\citet{fuest2018scalar}). \citet{wang2007dissipation} introduced a cut-off length scale ($\lambda_{\beta})$ (see also \citet{wang2008spatial}) as the inverse of the wavenumber at which 2$\%$ peak dissipation spectrum occurs. \citet{wang2007dissipation} found that, when normalized by $\lambda_{\beta}$, the dissipation spectra of temperature and mixture fraction nearly collapsed. 
%However, due to an increased noise level in the measured data, there is no evidence of the behaviour in higher wavenumber ranges. 
%
%Vaishnavi et al., \citet{vaishnavi2008spatial} used the Batchelor scale to normalize the passive scalar dissipation spectra. Using the result of DNS of spatial jet flame, they found that the normalized scalar dissipation rate has a peak around 0.26 times Batchlor scale. It is interesting to remember that the model TKE dissipation of Pope for constant density incompressible non-reactive flows (see Eq.~\ref{eq:turb:ModelDissSpectrum3D}) also has a peak around 0.26 times Kolmogorov's length scale.
Unlike non-reactive flows, the studies on the velocity and also the scalars spectra in reactive flows using DNS are limited in the literature. \citet{Knaus2009} investigated the effect of heat release on the velocity and scalars (mixture fraction and temperature) spectra obtained by DNS of temporally evolving reacting shear layers. Differential diffusion was not considered and effects of flame dynamics like extinction/re-ignition were neglected by using fast chemistry approach. Surprisingly, it was found that the effect of heat release can be well scaled out by using Favre averaged turbulence quantities in the velocity and mixture fraction spectra. \citet{Kolla2014} studied the energy spectra in premixed flames using DNS of temporally evolving jets. In agreement with \citet{Knaus2009}, they found that, in the inertial range, the classical scaling laws using Favre averaged quantities are applicable. Moreover, they saw that in high wavenumbers, the laminar flame thickness  (${\delta_{L} }$) produces a better collapse while disrupts the collapse in the inertial range. \citet{Kolla2016}, in the DNS study of the dissipation spectra of premixed jets, observed that scalars spectra collapse when normalized by their corresponding Favre mean dissipation rate and $\lambda_{\beta}$. However, in contrast to \citet{wang2007dissipation} and \citet{Knaus2009}, they saw that the normalized dissipation spectra in all the cases deviate noticeably from those predicted by the classical scaling laws for constant-density turbulent flows. It seems that the Favre form of Kolmogorov scaling is able to collapse the spectra computed in different conditions, both in non-premixed and premixed jet flames, in the inertial range. Its applicability in the near dissipation and far dissipation ranges is still controversial.     

The focus of this work is to study the velocity and its dissipation spectra in both inertial and near dissipation ranges of turbulent non-premixed jets using DNS databases. Differential diffusion has been considered in the DNS and the flames experience high levels of local extinction. 
%The work follows that of \citet{Knaus2009}, however, the effect of the differential diffusion is also accounted for. Further, \citet{Knaus2009} used the flame sheet approximation which assumes that the chemical reactions occur on much shorter time scales than the smallest flow scales. Using the imposed one-way coupling between chemistry and turbulence, they eliminated the effect of flame dynamics in  extinction/re-ignition regimes. The DNS cases used in the current study are all experiencing such transient effects. In the current work, the scaling of the spectra will be studied in such conditions to see if a consistency exists with the results of \citet{Knaus2009}. 
The objective is to study the Favre scaling laws proposed by \citet{Knaus2009} and to compare the spectra with the model spectrum of Pope (\citet{Pope2000turbulent}). After observing the collapse of spectra extracted from different flame positions and regimes,  Pope's model spectrum (both energy and dissipation) will be analysed and adapted for reactive DNS databases.

\section{DNS databases and the velocity spectra}\label{sec:appriori:DNS}
The DNS databases that are used in the current study are well documented DNS of temporally evolving jet flames with relatively detailed kinetics, performed in Sandia National Laboratories by S3D code (\cite{Hawkes2007}). Two flames with different bulk jet Reynolds numbers, namely cases M ($\Rey=4979$) and H ($\Rey=9079$) have been selected. The setup is a double shear layer, with a fuel (mixture of $\mathrm{CO}, \mathrm{H_2}, and \mathrm{N_2}$) stream in the middle surrounded by two counter-flowing oxidizer (air) streams. The fuel jets' initial widths are $H=0.96$ and $1.37 mm$ for cases M and H, respectively. The Reynolds number is defined as $\Rey \equiv UH/\nu_{fuel}$, with $U=194m/s$ and $276m/s$ the difference between the fuel and the oxidizer streams velocities for cases M and H, respectively. $\nu_{fuel}$ is the initial kinematic viscosity of the fuel. More details about the setup can be found in \cite{Hawkes2007}. 
The flames experience high levels of local extinctions followed by re-ignition. At $t=20t_j$, where $t_j=H/{U}=5\mu s$ is the ``transient jet time'', both flames experience maximum levels of local extinction. At other selected instants, i.e. $t=30t_j$, and $t=35t_j$, the flames are in re-ignition mode. 
Since the computational configuration has two periodic boundary conditions in $Ox$ (stream-wise) and $Oz$ (span-wise) directions, the flame is statistically 1D (in crosswise, $Oy$ direction) and the desired statistics can be obtained at an instant of time using the data in each statistically homogeneous $Oxz$ planes at each $Oy$. As an example, the streamwise mean velocity can be defined as $
\underline{u_1}(y,t)=({\sum_{n3=1}^{n3=N_z}\sum_{n1=1}^{n1=N_x}u_1(x_{n1},y,z_{n3},t)})/({N_xN_z})$, where $N_x$ and $N_z$ are the number of computational cells in $Ox$ and $Oz$ directions respectively. Then the Favre fluctuation of streamwise velocity is $u_1^{\prime\prime}=u_1-\underline{u_1}_f$ with $\underline{u_1}_f=\underline{\rho u_1}/\underline{\rho}$, the Favre averaged velocity. 1D velocity spectra  in streamwise direction ($E_{11}^{1D}(\kappa_{x})$ with $\kappa_x$ the longitudinal component of the wavenumber vector) can be calculated by 1D discrete Fourier transforms of all $u_1^{\prime\prime}$ data points in $Ox$ direction (768 and 864 in cases M and H, respectively), and averaging using all ensembles of equal $\kappa_x$ on $Oz$ (512 and 576 in cases M and H, respectively). In this way, 1D spectra will be constructed on each statistically homogeneous $Oxz$ plane. Uniform meshes have been used for both DNS cases with $(N_x,N_y,N_z)$ equals to $(768,896,512) and (864,1008,576)$ for cases M and H, respectively.
Seven different planes across the double shear layers are selected to compare the spectra. The planes are $P_0$: the $Oxz$ plane at $y=0$, i.e., the central plane, $P_1-P_4$, the planes corresponding to the maximum density and temperature variances, maximum Favre average $\mathrm{OH}$ mass fraction and maximum TKE, respectively and further, $P_5-P_6$ the planes corresponding to the Favre mean mixture fraction ($\underline{Z}_f$) equals to 0.7, and 0.422 (stoichiometric mixture fraction), respectively.  
 
\section{Normalized Spectra Using Favre Averaged Kolmogorov's Scales}\label{sec:appriori:NormalSpectra}
\begin{figure}
\centering
	\subfigure[]{\label{fig:appriori:E111DDifferentTimesPlanesHcase.a}\includegraphics[width=0.3\textwidth]{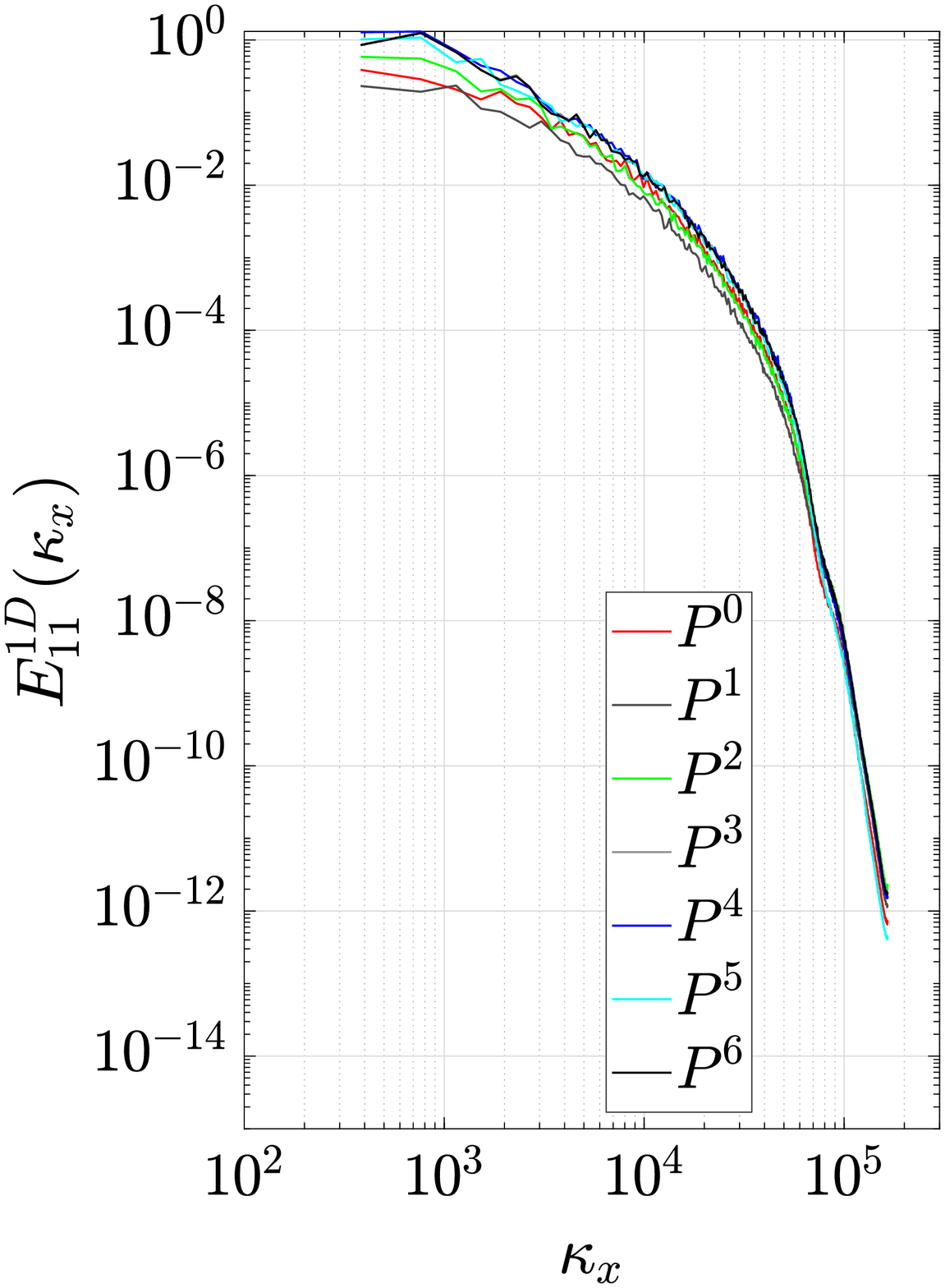}}
	\subfigure[]{\label{fig:appriori:E111DDifferentTimesPlanesHcase.b}\includegraphics[width=0.3\textwidth]{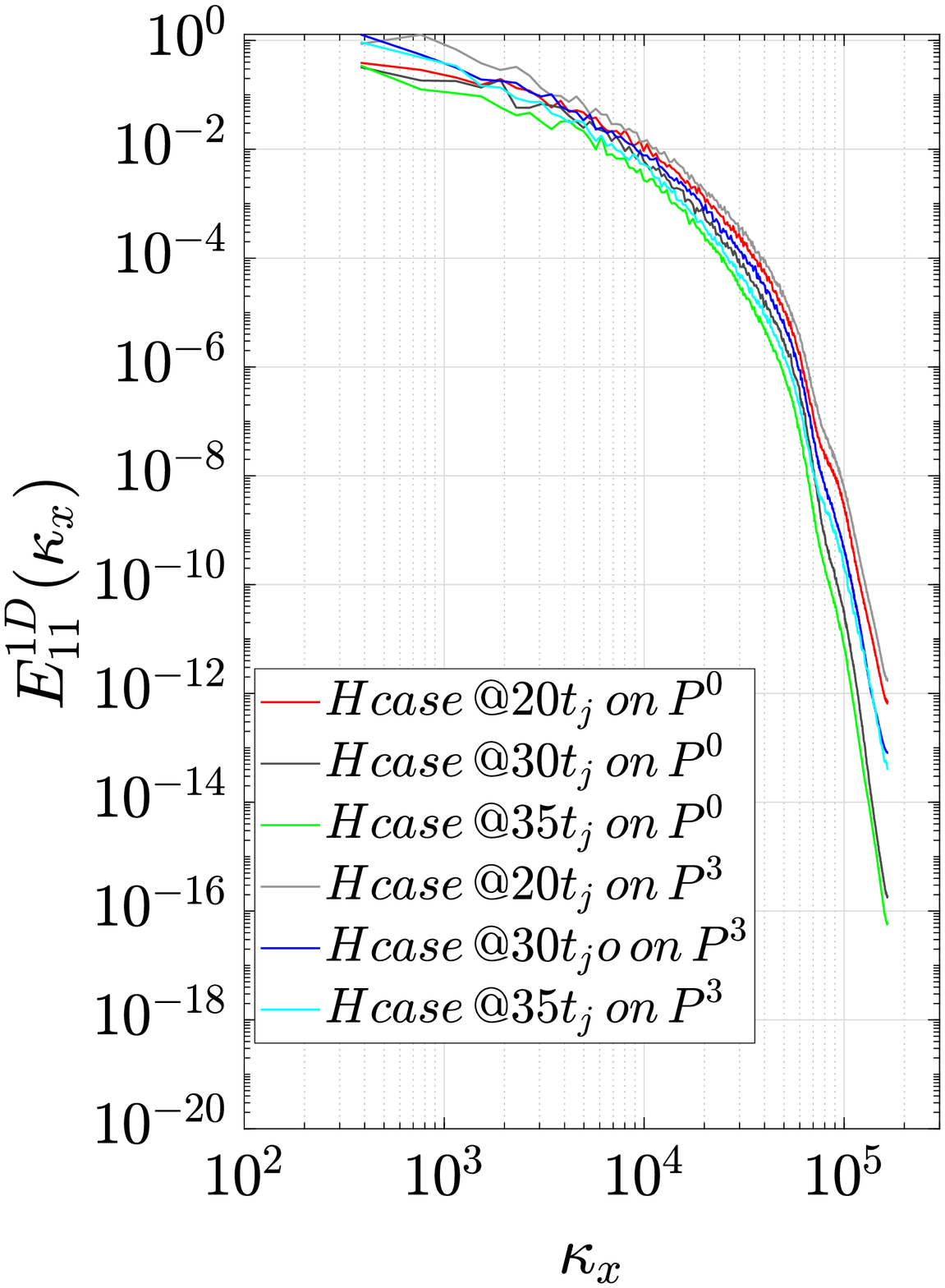}}
	\subfigure[]{\label{fig:appriori:E111DDifferentTimesPlanesHcase.c}\includegraphics[width=0.3\textwidth]{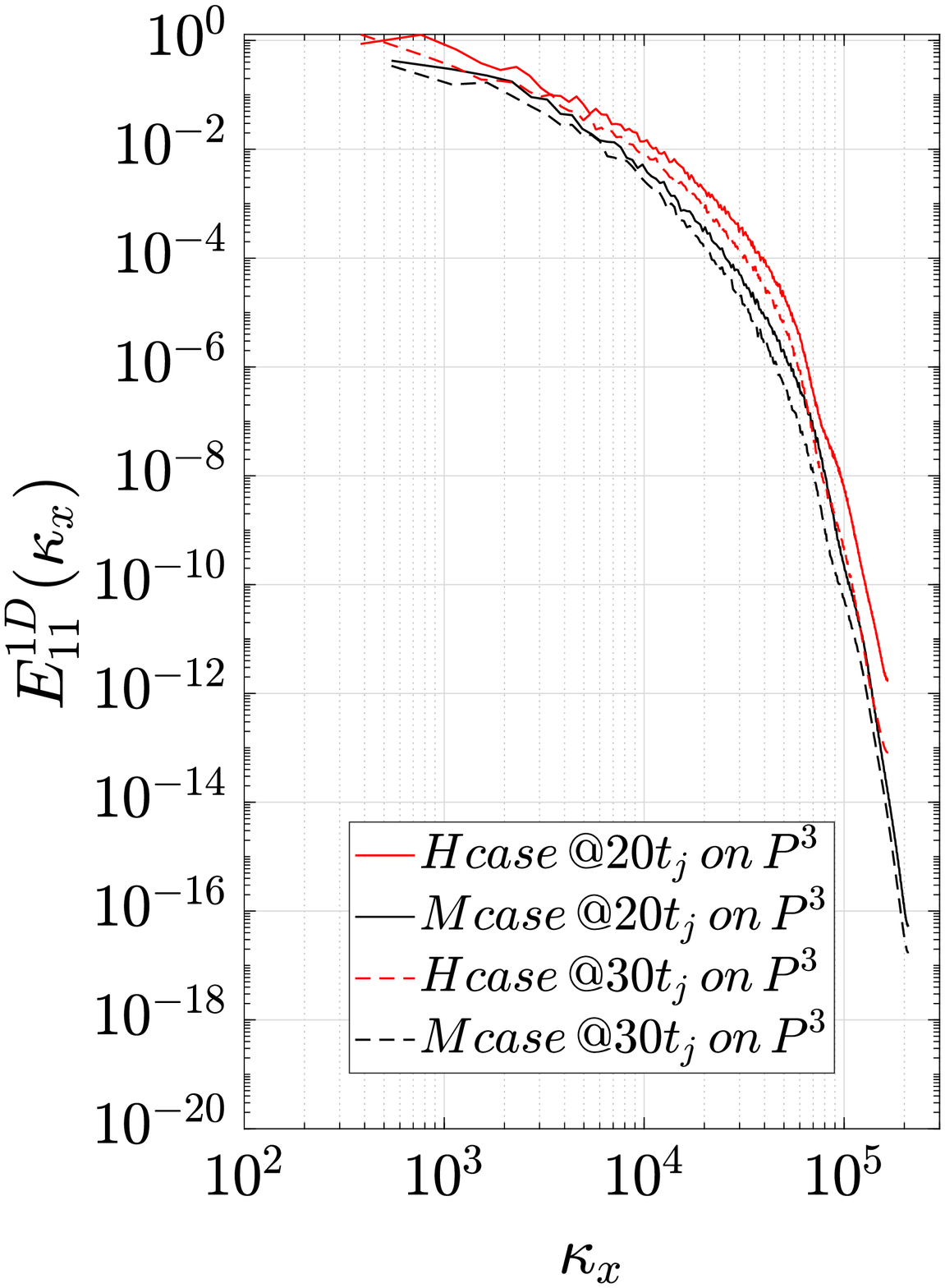}}\\
	\subfigure[]{\label{fig:appriori:E111DDifferentTimesPlanesHcase.d}\includegraphics[width=0.3\textwidth]{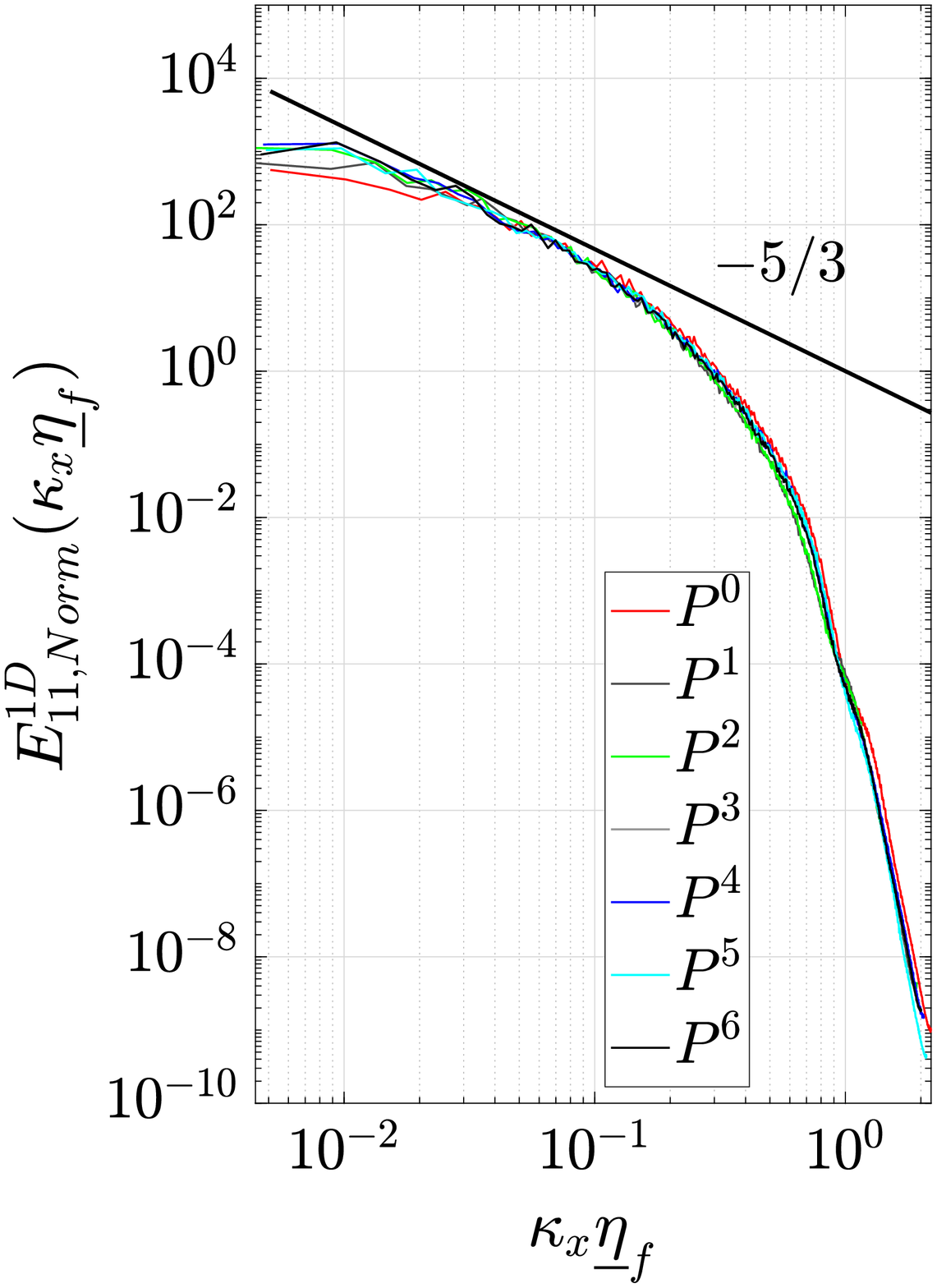}}
	\subfigure[]{\label{fig:appriori:E111DDifferentTimesPlanesHcase.e}\includegraphics[width=0.3\textwidth]{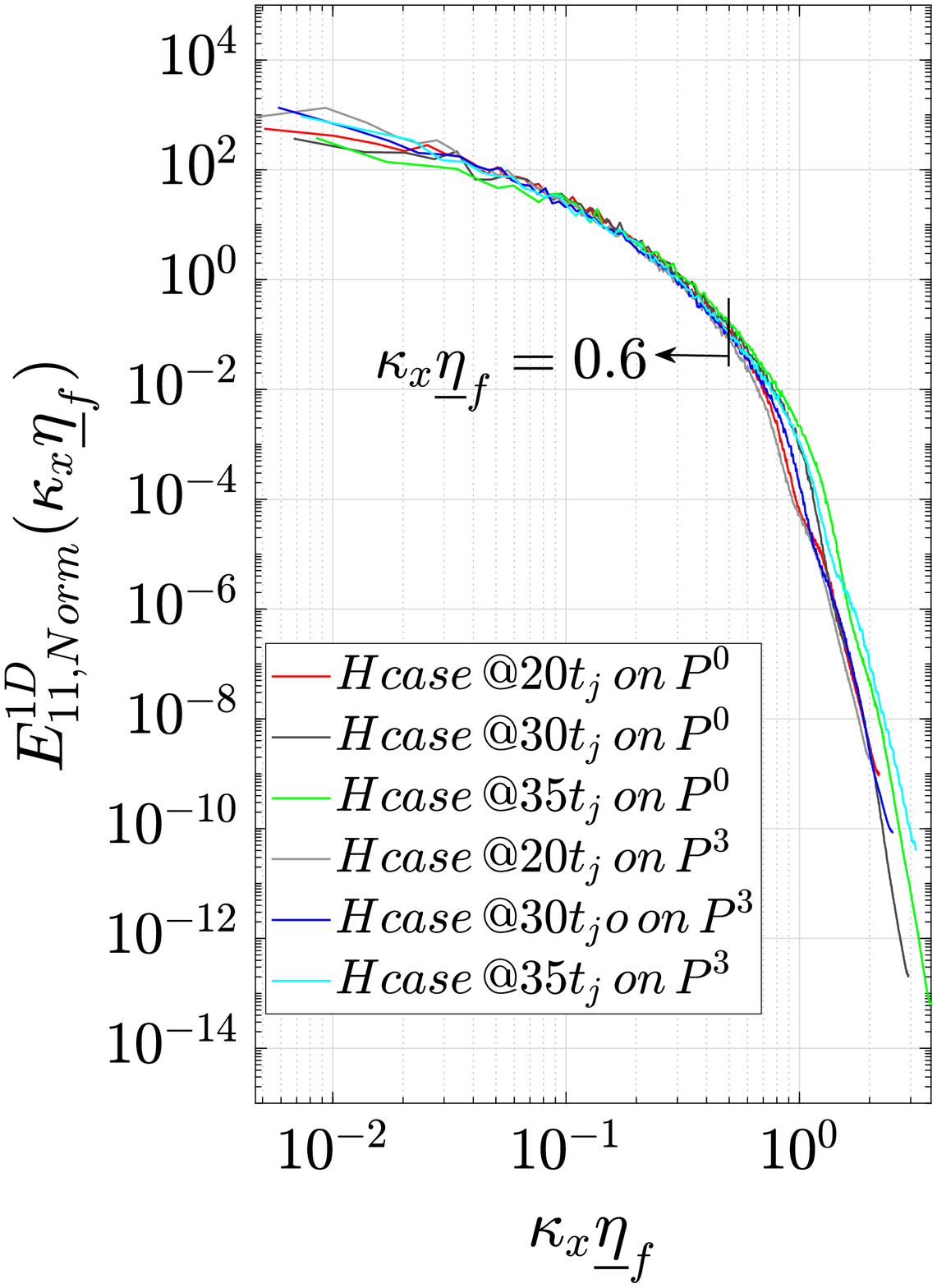}}
	\subfigure[]{\label{fig:appriori:E111DDifferentTimesPlanesHcase.f}\includegraphics[width=0.3\textwidth]{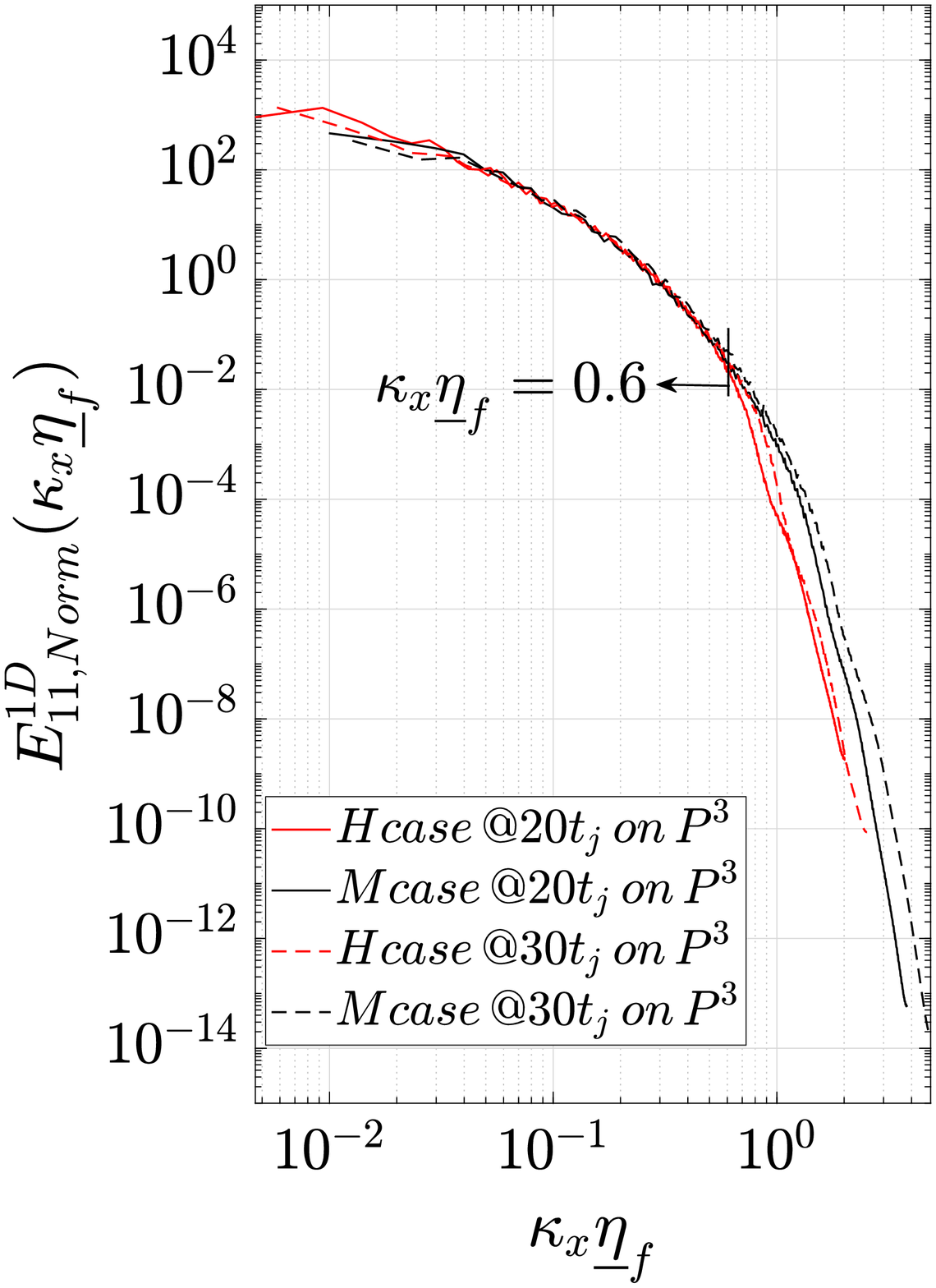}}		
	\caption{\label{fig:appriori:E111DDifferentTimesPlanesHcase}1D longitudinal velocity spectra across the reacting shear layer. (\textbf{Top}) non-normalized, (\textbf{bottom}) normalized using the Favre averaged $\eta$ and $\varepsilon$. Spectra for (\textbf{left}) case H on different cross planes at $t=20t_j$, (\textbf{middle}) for case H on the central and $\underline{Z}_f=Z_{st}$ plane and  at  $t=20, 30, 35t_j$, (\textbf{right}) cases M and H at $t=20, 30t_j$ on $\underline{Z}_f=Z_{st}$ plane.}
\end{figure}

\citet{Knaus2009} studied 1D energy spectra using DNS of temporally evolving reacting shear layers (methane-air and hydrogen-air) with a one step global reaction. The 1D energy spectrum was computed using 1D velocity spectra, viz. $E^{1D}(\kappa_x)=E_{11}^{1D}(\kappa_{x})+E_{22}^{1D}(\kappa_{x})+E_{33}^{1D}(\kappa_{x})$.
%where $E_{ii}^{1D}(\kappa_{x})$ can be computed by the 1D Fourier transforms in 
 In this work, however, it is preferred to work with the individual components of $E^{1D}$, e.g. $E_{11}^{1D}$. 
%This is because in the literature there is no theoretical background for the 1D energy spectrum. The Kolmogorov spectrum is a 3D energy spectrum and equations were derived to relate it to the 1D spectra (i.e., components of Eq.~\ref{eq:appriori:EkKolmo1Dxx} not the sum). A universal Favre normalized 1D velocity spectrum for reactive flows can be defined as:
This is due to the ease of conversion to 3D spectra using existing theories. 
%because using theoretical formula with isotropic assumptions, it can be easily converted to 3D energy spectrum. This will be discussed later. 
 The 1D streamwise velocity spectra for case H are presented in the first row of figure~\ref{fig:appriori:E111DDifferentTimesPlanesHcase}. It is obvious that the non-normalized spectra do not collapse on a single curve in the low-medium range of $\kappa_{x}$. 
 A Favre normalized 1D velocity spectrum is defined as:
 \begin{equation}\label{eq:appriori:EkKolmo1DxxNormal}
 E_{11,Norm}^{1D}(\kappa_x \underline{\eta}_f) \equiv \frac{E_{11}^{1D}(\kappa_x)}{{\underline{\varepsilon}_f}^{2/3} {\underline{\eta}_f}^{5/3}}
 \end{equation}
 where the Favre averaged TKE dissipation rate, $\underline{\varepsilon}_f$, is exact and extracted from the DNS. 
 The normalization proposed in \citet{Knaus2009} is shown in the second row
%The plot shows the normalized 1D longitudinal velocity spectra versus the longitudinal wavenumber normalized by the Favre averaged Kolmogorov length scale (i.e. $\kappa_{x}\underline{\eta}_f$). 
%The Favre averaged TKE dissipation rate, $\underline{\varepsilon}_f$ is exact and extracted from the DNS using Eq.~\ref{eq:turb:exactDissTKE} in Sec.~\ref{sec:turb:FANS}. 
where the black line shows $-5/3$ slope. It can be observed that the $\kappa^{-5/3}$ scaling exists in agreement with Kolmogorov's hypothesis although it is not extended over a wide range due to the relatively low Reynolds number of the jets. The collapse of the spectra (i.e., existence of the universal curve), is in agreement with \citet{Knaus2009} and \citet{Kolla2014}, and exists  in the whole wavenumber range except the very low wavenumber in which the effect of anisotropy exists. 
In figures ~\ref{fig:appriori:E111DDifferentTimesPlanesHcase.b} and ~\ref{fig:appriori:E111DDifferentTimesPlanesHcase.e},   
the non-normalized and normalized 1D velocity spectra on $P^{0}$ and $P^3$ for case H at different time instants corresponding to different flame dynamics are shown, respectively. Moreover, the spectra  on $P^3$ for cases with different Re are depicted in figures ~\ref{fig:appriori:E111DDifferentTimesPlanesHcase.c} and ~\ref{fig:appriori:E111DDifferentTimesPlanesHcase.f}. 
%In the middle plots of figure~\ref{fig:appriori:E111DDifferentTimesPlanesHcase} the non-normalized and normalized 1D velocity spectra on $P^{0}$ and $P^3$ are plotted for case H at different time instants corresponding to different flame dynamics. 
%(the central $Oxz$ plane at $y=0$, $P^{0}$ and the $Oxz$ plane corresponding to maximum Favre mean $\mathrm{OH}$ mass fraction, $P^3$). 
%Next in the left plots of figures~\ref{fig:appriori:E111DDifferentTimesPlanesHcase}  the non-normalized and normalized velocity spectra on $P^3$ are plotted for cases with different Re (cases M and H).
Since for each case it was already observed that the collapse of the spectra extracted from different planes is acceptable (case M not shown here for the sake of brevity), in the plots on the right side of the figure only the results on $P^3$ are plotted.  The collapse of the spectra for case H starts to deteriorate after $\kappa_x\underline{\eta}_f\approx0.6$. However, for case M the range of wavenumbers where a good collapse is observed is wider, i.e. $\kappa_x\underline{\eta}_f\approx1$ (see figure.~\ref{fig:appriori:E111DDifferentTimesPlanesHcase.f}). 
An inflection is observed in all plots of figure~\ref{fig:appriori:E111DDifferentTimesPlanesHcase}. The normalized wavenumber corresponding to the inflection point is different between the two DNS cases. In \citet{Kolla2014}, using DNS of temporally evolving premixed jet flames simulated by the same S3D code, a similar inflection was observed. In that paper, it was hypothesized that the inflection is due to the pressure-velocity coupling at the laminar flame scales. This point will be better described later, but for the moment, it is good to remember that a change of slope and the following breakdown of the collapse of the spectra is observed before the inflection.  
%
%Next we will compare the collapse of the spectra for cases with different Re (cases M and H) and in different regimes. Since for each case it was already observed that the collapse is acceptable, here only the results on the plane corresponding to the maximum Favre mean $\mathrm{OH}$ mass fraction (i.e., $P^3$) is plotted 
%in figure~\ref{fig:appriori:E111DDifferentTimesPlanesHcase.c} and \ref{fig:appriori:E111DDifferentTimesPlanesHcase.f}. The collapse of the spectra extracted from two different cases with two different Re and physically two different levels of extinction using the normalization method is acceptable. 
%It is seen that the normalized spectra from extinction regime and re-ignition regime, although differ a lot (figure~\ref{fig:appriori:E111DDifferentTimesPlanesHcase.a}), collapse well using $\underline{\eta}_f$ as a length scale (figure~\ref{fig:appriori:E111DDifferentTimesPlanesHcase.b}) in the inertial range and part of near dissipation range (up to around $\kappa_x\underline{\eta}_f\leq 0.6$). It can be clearly seen that the collapse is improved mainly in near dissipation range using the cut-off scale $\lambda_{\beta}$ (figure~\ref{fig:appriori:E111DDifferentTimesPlanesHcase.c}).

\begin{figure}
\centering
\subfigure[]{\label{fig:appriori:E11-compensatedWithSlopes-at0099Hcase.a}\includegraphics[width=0.35\textwidth]{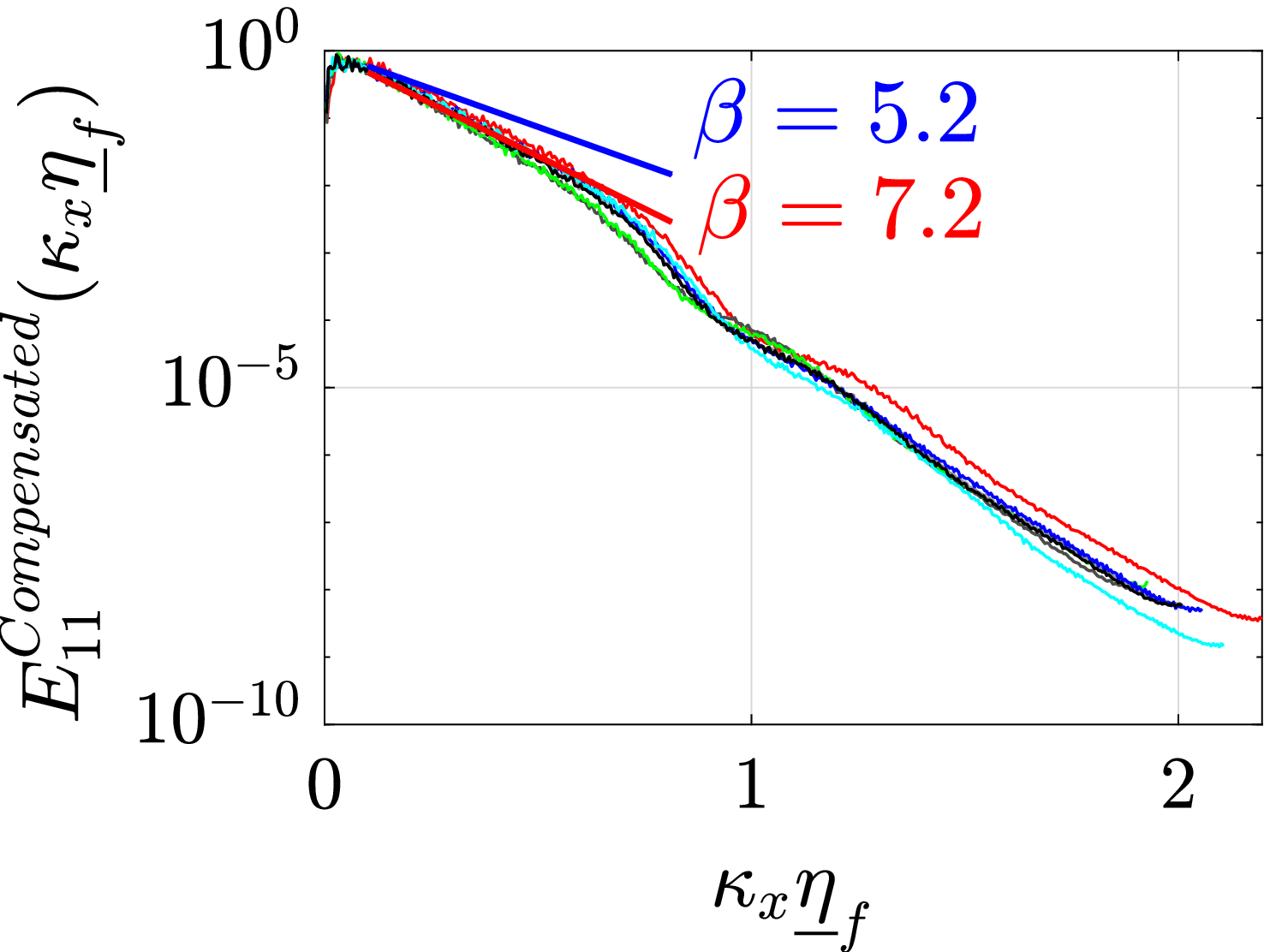}}
	\subfigure[]{\label{fig:appriori:E11-compensatedWithSlopes-at0099Hcase.b}\includegraphics[width=0.35\textwidth]{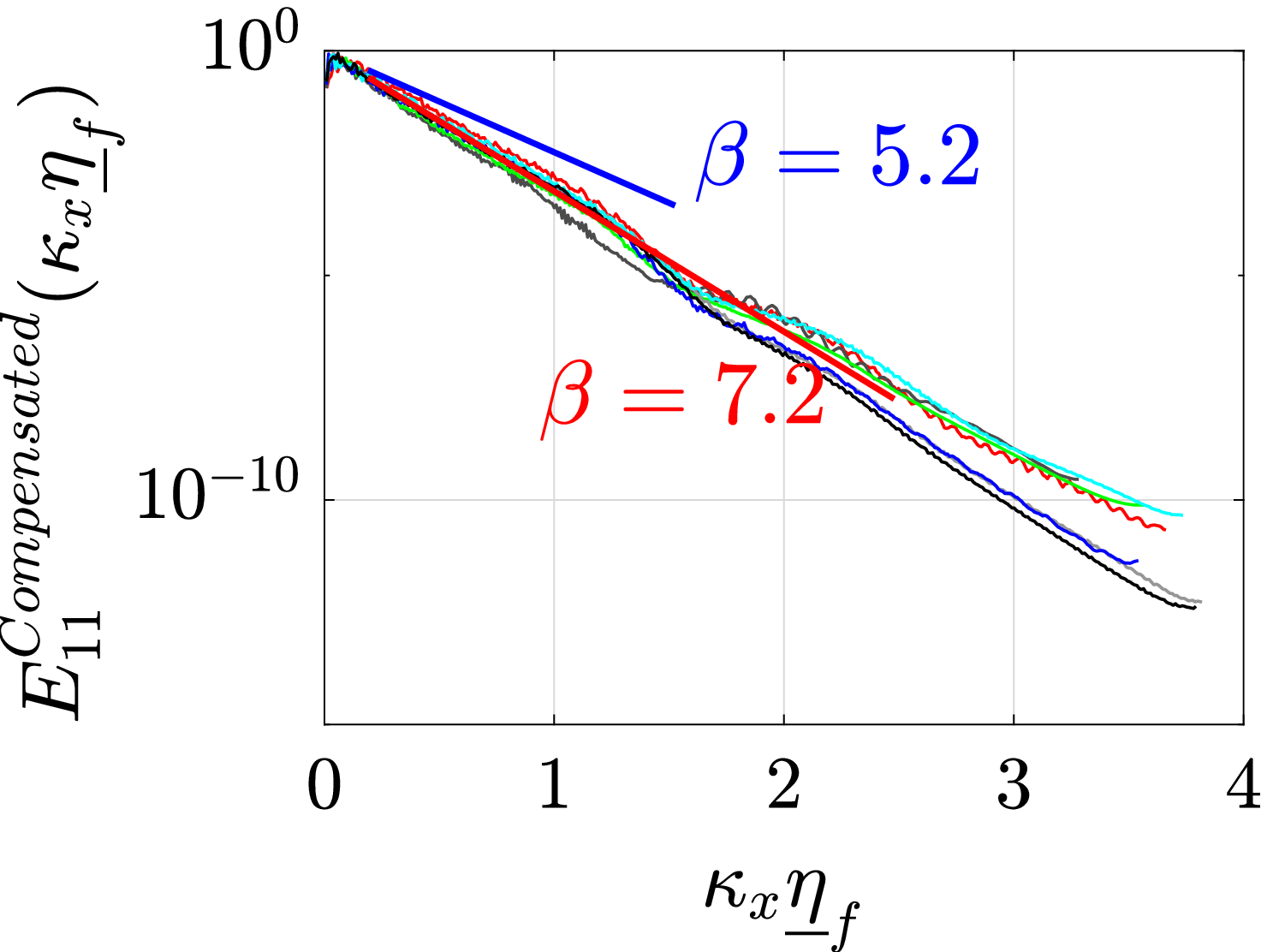}}
	\subfigure[]{\label{fig:appriori:E11-compensatedWithSlopes-at0099Hcase.c}\includegraphics[width=0.35\textwidth]{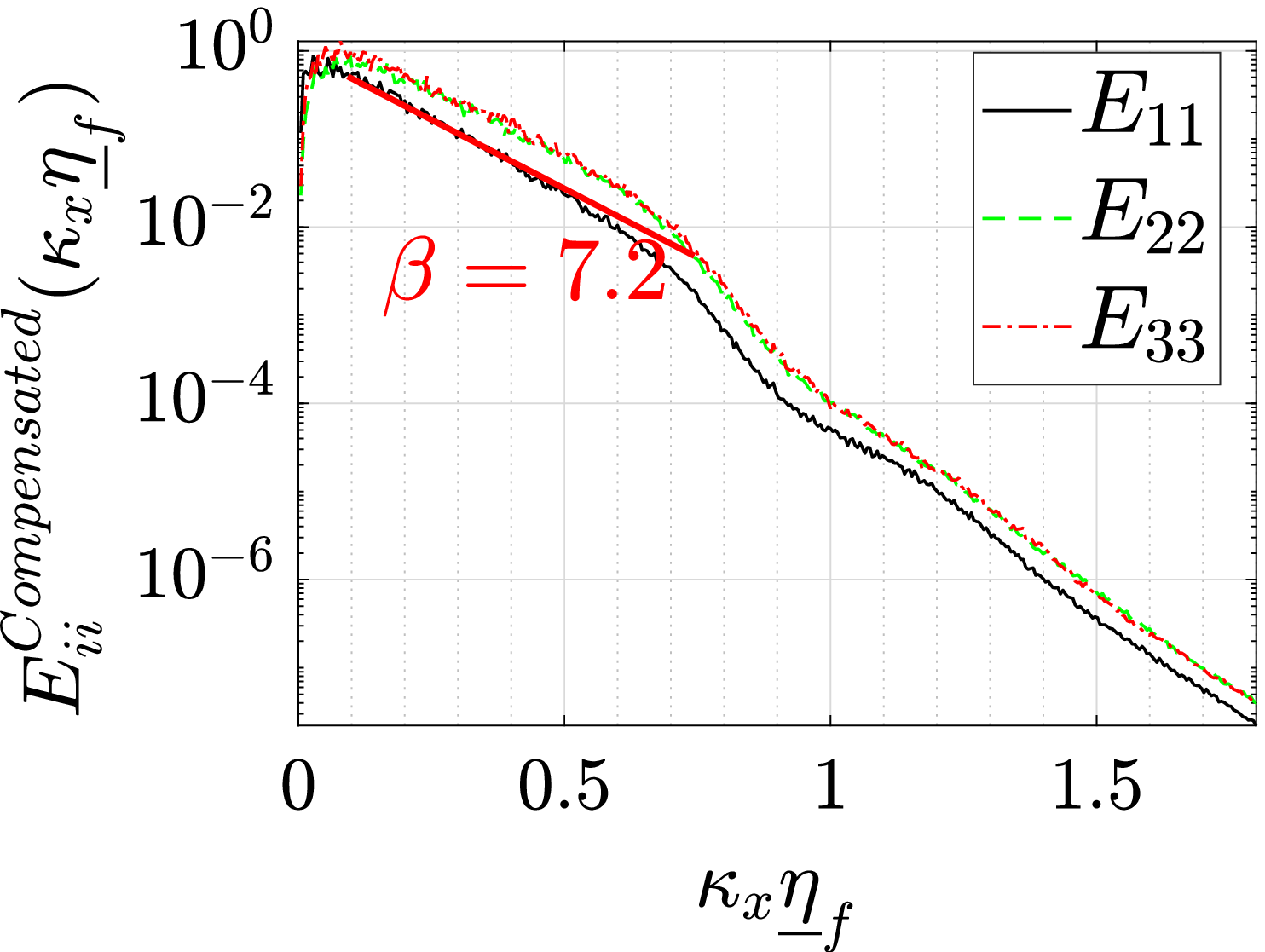}}
	\caption{1D compensated longitudinal spectra  at $t=20t_j$ for (\textbf{left}) case H, and (\textbf{middle}) case M extracted from different planes across the jets, and (\textbf{right}) case H extracted from the Favre mean Stoichiometric mixture fraction plane. In (a) and (b), the same colors as in figure~\ref{fig:appriori:E111DDifferentTimesPlanesHcase} has been used.}
	\label{fig:appriori:E11-compensatedWithSlopes-at0099Hcase}       % Give a unique label
\end{figure}

%\begin{figure*}
%\centering
%	\begin{subfigure}{0.3\textwidth}
%	\includegraphics[width=\textwidth]{E11DVSkxat0099Hcase_incomp_corr.eps}
%	\caption{\label{fig:appriori:Ek1DVSkx_2at0099Hcase3.a}}
%	\end{subfigure}
%	\begin{subfigure}{0.3\textwidth}
%		\includegraphics[width=\textwidth]{E111DNormalKolmoVSkxat0099Hcase_incomp_corr.eps}
%		\caption{\label{fig:appriori:Ek1DVSkx_2at0099Hcase3.b}}
%	\end{subfigure}
%	\caption{1D longitudinal spectra on different planes across the reacting shear layer at $20t_j$ for case H, (\textbf{a}) Non-normalized; (\textbf{b}) Normalized using Favre averaged $\eta$ and $\varepsilon$.}
%	\label{fig:appriori:E11DVSkx_2at0099Hcase3}%
%\end{figure*}	
%\begin{figure*}
%\centering
%	\begin{subfigure}{0.3\textwidth}
%	\includegraphics[width=\textwidth]{Ek1DVSkx58-compensated-at0099Hcase_incomp_corr.png}
%		\caption{\label{fig:appriori:Ek1DVSkx3-compensated-at0099Hcase.a}}%
%	\end{subfigure}
%	\begin{subfigure}{0.3\textwidth}
%	\includegraphics[width=\textwidth]{E1133-compensated-at0099Hcase_incomp_corr.png}
%		\caption{\label{fig:appriori:Ek1DVSkx3-compensated-at0099Hcase.b}}%
%	\end{subfigure}
%	\caption {1D compensated spectra on different planes across the reacting shear layer at maximum extinction time, $20t_j$ extracted from H case DNS, (\textbf{a}) 1D compensated energy spectra; (\textbf{b}) 1D compensated longitudinal spectra.}
%		\label{fig:appriori:Ek1DVSkx3-compensated-at0099Hcase}%
%\end{figure*}
In near dissipation range, \citet{Kraichnan1959}, for non-reactive flows,  proposed that the 3D normalized energy spectrum $E_{Norm}$ has an exponential drop-off of the form 
%\footnote{The theory is for non-reacting incompressible flows so the correct form is to use $\underline{\eta}$ rather than $\underline{\eta}_f$, however, since we are dealing with the reactive flow, we adopt the Favre average quantities in all formula}
% \begin{equation}\label{eq:appriori:KraichnanModel}
 $E_{Norm}(\kappa\underline{\eta}_f)=A\left(\kappa\underline{\eta}_f\right)^\alpha \exp \left(-\beta\left(\kappa\underline{\eta}_f\right)^n \right)$.
% \end{equation} 
 The value of $n=1$ is supported by two-point closure theories (\citet{ishihara2005energy}) for $\kappa\underline{\eta}_f\gg1$. For near dissipation range (say $0.5\leq\kappa\underline{\eta}_f\leq1.5$) this value is supported by experiments 
%(e.g., \citet{Sreenivasan1985,saddoughi1994local})  
(e.g., \citet{saddoughi1994local})  
%and DNS e.g., \citet{martinez1997energy,Kaneda2006,ishihara2005energy}. 
and DNS (e.g., \citet{Kaneda2006}). 
 If the compensated form of spectrum (i.e., $E_{Norm}(\kappa\underline{\eta}_f)\times(\kappa\underline{\eta}_f)^{5/3}$) is plotted against $\kappa\underline{\eta}_f$ in a log-linear plot and if a straight line is observed, it can be concluded that $\alpha=-5/3$ and the slope of the line gives $\beta$. It should be mentioned that we expect a similar functional form of the spectrum (with different $A$) when analysing the 1D velocity spectrum rather than the 3D energy spectrum. This is shown in figure~\ref{fig:appriori:E11-compensatedWithSlopes-at0099Hcase}. For example for case H (figure~\ref{fig:appriori:E11-compensatedWithSlopes-at0099Hcase.a}) in the range of  $0.1\leq\kappa_x\underline{\eta}_f\leq0.6$ the straight line is observed. 
% The observation of the straight line implies that $\alpha=-5/3$, consistent with Kolmogorov's spectrum. This value of  $\alpha$ was also reported in the previous studies of non-reactive cases \citet{saddoughi1994local}. 
 However, unlike the well documented slope of $\beta=5.2$ for non-reactive flows (\citet{saddoughi1994local}), in the near dissipation range, the observed slope is $\beta\approx7.2$. The same slope can be observed for lower Re case M (see  figure~\ref{fig:appriori:E11-compensatedWithSlopes-at0099Hcase.b}). 
 The inflection observed before in figure~\ref{fig:appriori:E111DDifferentTimesPlanesHcase} is more evident  near $\kappa_x\underline{\eta}_f\approx0.8$ for case H in  figure~\ref{fig:appriori:E11-compensatedWithSlopes-at0099Hcase.a}. Before and after that point, there is a change in the slope. 
% to approximately  $\beta\approxeq10$. 
The change of the slope starts from $\kappa_x\underline{\eta}_f\approx 0.6$ in case H.
 A change of the slope was previously observed in non-reactive flows however in a very far dissipation range around $\kappa\underline{\eta}\approx4$ in the DNS of isotropic turbulence with ${Re_{\lambda}}$ range similar to the cases considered here (\citet{martinez1997energy}). The spectra computed for case M, which has a higher resolution show the extended range with $\beta=7.2$ slope and the transition occurs at higher normalized wavenumbers, $\kappa_x\underline{\eta}_f>1.4$ (see  figure~\ref{fig:appriori:E11-compensatedWithSlopes-at0099Hcase.b}). 
 Different components of the 1D energy spectrum are depicted in  figure~\ref{fig:appriori:E11-compensatedWithSlopes-at0099Hcase.c} for case H at $t=20t_j$. 
 %figure~\ref{fig:appriori:Eii-compensatedWithSlopes-at0099Hcase.a} shows the longitudinal ($E_{11}^{Compensated}(\kappa_x\underline{\eta}_f)$) and the transverse ($E_{22}^{Compensated}(\kappa_x\underline{\eta}_f)$ and $E_{33}^{Compensated}(\kappa_x\underline{\eta}_f)$) spectra extracted from the mean stoichiometric plane of case H at $t=20t_j$. figure~\ref{fig:appriori:Eii-compensatedWithSlopes-at0099Hcase.b} shows the same in a different plane which is the central plane. 
 It is observed that the behaviour of the spectra are consistent with the theory, i.e. $E_{22}\approx E_{33}>E_{11}$, before and after inflection. So it is not clear that if the inflection and change of slope is an intrinsic flame behaviour like what hypothesized in \citet{Kolla2014} for premixed flames (pressure-velocity coupling at laminar flame scales) or is the effect of a tenth order filter which was used in the DNS code. This needs more future investigations. So the focus of our study will be only on the wavenumbers before inflection where the well behaved straight line and the collapse is observed.
 
 The compensated form of the velocity spectra can be used to detect the Kolmogorov constant ${{C_{K}}}$ (\citet{saddoughi1994local}). \citet{Sreenivasan1995}, examined a wide range of experiments for non-reactive flows and suggested a range of  $0.53\pm0.055$ for the 1D Kolmogorov constant. When multiplied by ${55}/{18}$ (see \citet{Monin1971statistical}), it can be transformed to the constant of the 3D spectrum. In this way the value for 3D Kolmogorov's constant is $1.62\pm0.17$. Approximately the same value was observed in  non-reactive DNS studies (see e.g., 
 %  \citet{Yeung1997,Ishihara2009}. 
 \citet{Ishihara2009}). 
 % between $E_{11}^{1D}$ and $E^{3D}$ , $E_{11}^{1D}(\kappa_x)= {18}/{55} C_K (\underline{\varepsilon})^{2/3}\kappa_x^{-5/3}$, as $1.62\pm0.17$). Approximately the same value was found from non-reactive DNS studies, see e.g.
%  \citet{Yeung1997,Ishihara2009}. 
% \citet{Ishihara2009}. 
 \citet{donzis2010bottleneck} mentioned that it is difficult to find Kolmogorov's constant using the compensated form of spectra because a perfectly horizontal region does not exist. %This is related to the inertial range intermittency. 
 A closer inspection also shows that ${{C_{K}}}$ is not  a real constant and has a weak ${Re_{\lambda}}$ dependency 
% (see e.g. \citet{tsuji2009high,praskovsky1994measurements} 
 of about power 0.1 (see e.g. \citet{tsuji2009high}). It can be shown that it is due to the intermittency effects. Intermittency correction can be added to the $\kappa^{5/3}$ power law, however, 
 %it is believed that the tilted region can be seen only in high ${Re_{\lambda}}$ experiments or DNS (\citet{tsuji2009high}). 
 in low  ${Re_{\lambda}}$, the intermittency correction is very low, (order of $10^{-2}$).  
 %so in the current study we assume that intermittency does not have a huge effect on the TKE spectrum. 
 In the DNS of high ${Re_{\lambda}}$  ($Re_\lambda>700$) by  \citet{ishihara2016energy},  the compensated spectrum showed a very short nearly flat region, a tilted region, and a bump near $\kappa \underline{\eta}=0.1$. The bump was also observed in the experiments of  \citet{saddoughi1994local}.  This is called the bottleneck effect and has been discussed in detail by  \citet{donzis2010bottleneck}. Since in the current study the ${Re_{\lambda}}$ is relatively low, the two artefacts will be neglected and $C_K$ will be extracted from the compensated spectra.  
% For the first issue, the so called intermittency correction is added to the $\kappa^{5/3}$ power law, however, it is believed that the tilted region can be seen only in high ${Re_{\lambda}}$ experiments or DNS \citet{tsuji2009high}.
%  As we already mentioned, intermittency usually has a profound effect on the small scale scales (gradients) and high order statistics; the effect on the second order statistics (the spectrum in wavenumber space) is not high and can be considered to be negligible \citet{Yeung1997,Frisch1995turbulence,Sreenivasan1995}. 
  The value  of the 1D constant from $E_{11}^{1D}$ is found to be $\approx 0.75$ for the two flames. The value is higher than the reported values for non-reactive jets ($\approx 0.5$). %which (using isotropy assumption) corresponds to the value of $\approx 1.5$ for the 3D constant. 
%\begin{figure*}
%\centering
%	\begin{subfigure}{0.3\textwidth}
%	\includegraphics[width=\textwidth]{Eii-compensatedWithSlopesAtStoi-at0099Hcase_incomp_corr.eps}
%	\caption{\label{fig:appriori:Eii-compensatedWithSlopes-at0099Hcase.a}}
%	\end{subfigure} 
%		\begin{subfigure}{0.3\textwidth}
%	\includegraphics[width=\textwidth]{Eii-compensatedWithSlopesAtcenterPlane-at0099Hcase_incomp_corr.eps}
%	\caption{\label{fig:appriori:Eii-compensatedWithSlopes-at0099Hcase.b}}
%		\end{subfigure}
%	\caption{1D compensated spectra across the reacting shear layer at $t=20t_j$ for case H, (\textbf{a}) Extracted from the Favre mean Stoichiometric mixture fraction plane; (\textbf{b}) Extracted from the central plane.}
%	\label{fig:appriori:Eii-compensatedWithSlopes-at0099Hcase}%
%\end{figure*}	

\begin{figure}
\centering 
\subfigure{\label{fig:appriori:D111DVSkxat0099Hcase.a}\includegraphics[width=0.35\textwidth]{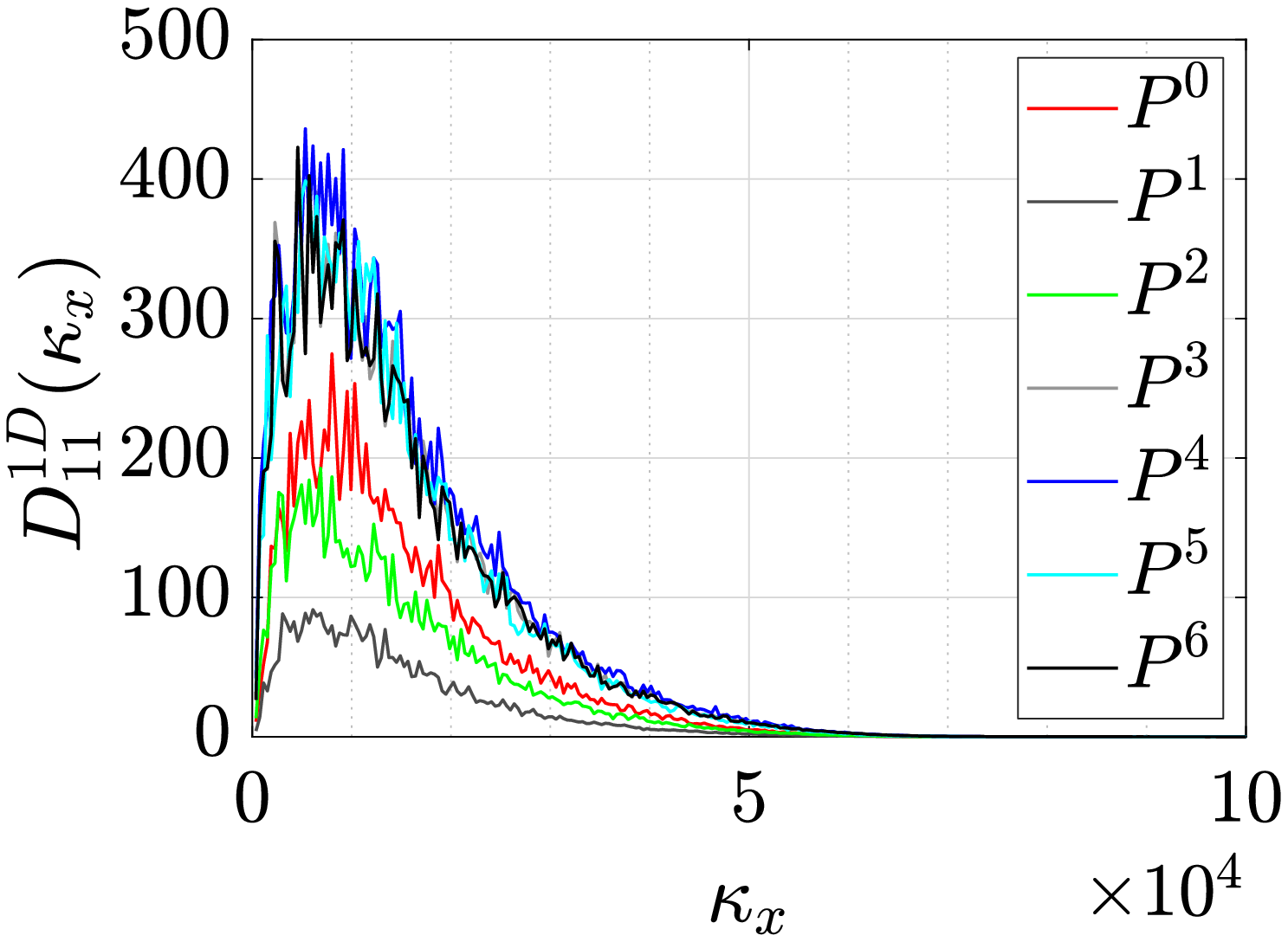}}
	\subfigure{\label{fig:appriori:D111DVSkxat0099Hcase.b}\includegraphics[width=0.35\textwidth]{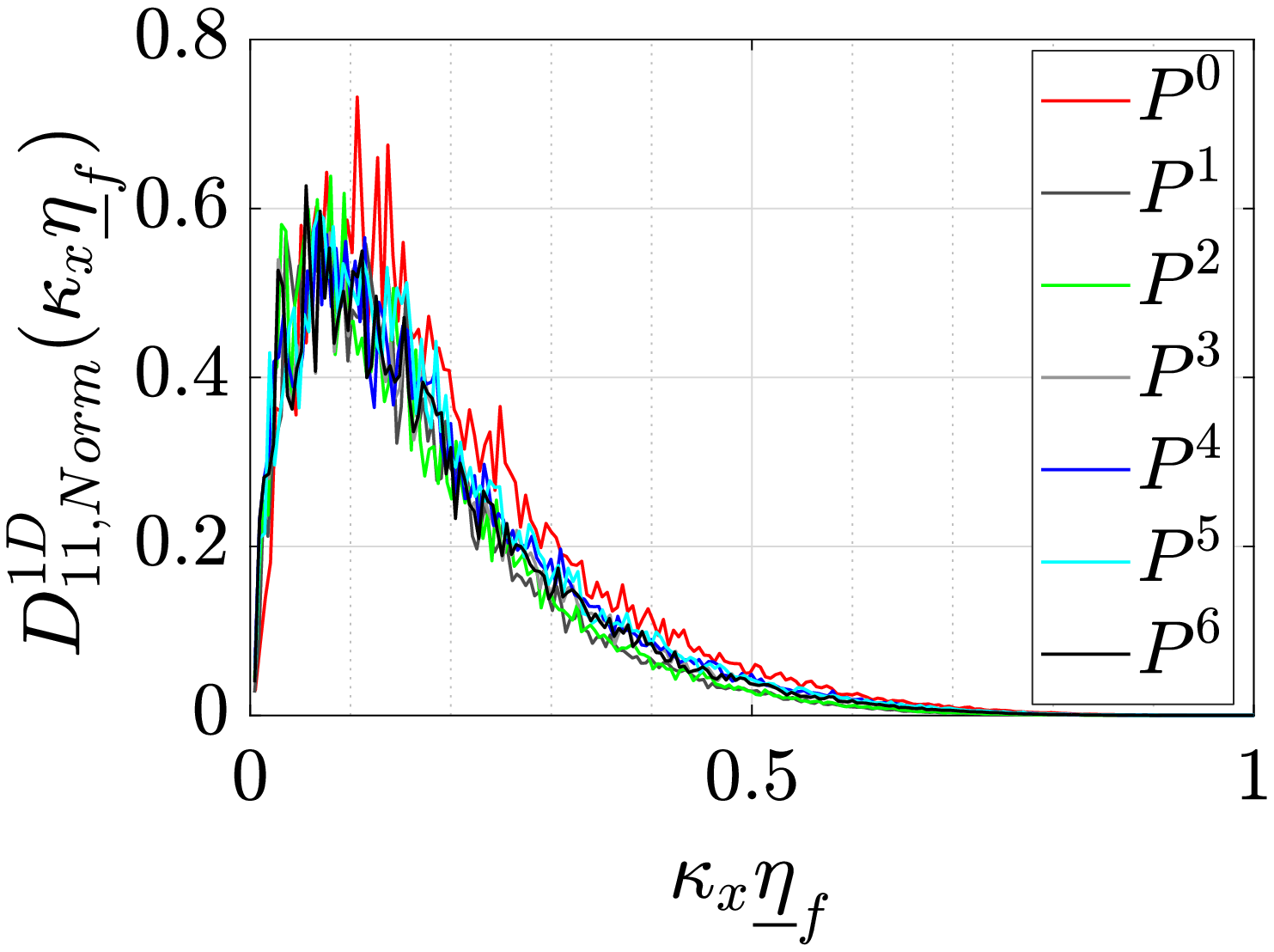}}
	\caption{1D longitudinal dissipation spectra across the reacting shear layer at $t=20t_j$ for case H, (\textbf{left}) Non-normalized spectra; (\textbf{right}) Normalized using Favre averaged $\eta$ and $\varepsilon$.}
	\label{fig:appriori:D111DVSkxat0099Hcase}%
\end{figure}

In figure~\ref{fig:appriori:D111DVSkxat0099Hcase}  the non-normalised and normalised 1D dissipation spectra are plotted. The normalised 1D dissipation spectrum reads:
\begin{equation}\label{eq:turb:DNormalKappaEtaDef}
D_{11,Normal}(\kappa_x \underline{\eta}_f) \equiv \dfrac{D_{11}(\kappa_x)}{\underline{u_\eta}^3}=\dfrac{D_{11}(\kappa_x)}{\underline{\eta}_f\underline{\varepsilon}_f},
 \end{equation}
where $D_{11}(\kappa_x)\equiv 2\underline{\nu}_f\kappa_x^2 E_{11}(\kappa_x)$  with $\underline{\nu}_f$ the Favre averaged kinematic viscosity. 
If one wanted to calculate the 3D TKE ($0.5 u_ {i}^{\prime\prime}u_ {i}^{\prime\prime}$) dissipation spectrum, i.e., $D^{3D}(\kappa)$, it was possible to directly transform the instantaneous dissipation rate, $\varepsilon$, to Fourier space. However, we keep using the conventional incompressible definition and evaluate the 1D dissipation of $u_1^{\prime\prime}u_1^{\prime\prime}$ instead of $D^{3D}(\kappa)$. 
It is observed in figure~\ref{fig:appriori:D111DVSkxat0099Hcase} that
by approaching the core jet region, the dissipation peak value increases. In the normalized dissipation spectra (the right plot in figure~\ref{fig:appriori:D111DVSkxat0099Hcase}), the peak location seems to collapse well and become approximately independent of ${Re_{\lambda}}$ . The peak occurs in $\kappa_x\underline{\eta}_f\approx 0.08$. This is of great importance. It is observed that the location of the peak of dissipation spectrum in non-premixed flames is lower than the non-reactive jet value and grid generated turbulence reported in \citet{Ganapathisubramani2008} and \citet{saddoughi1994local}, which is  $\sim0.1$. Further, the collapse of the peak location is acceptable when the dissipation spectra are normalized by $\underline{\eta}_f$ and $\underline{\varepsilon}_f$. 

%Moreover, we do not expect that the peak height of the normalized dissipation spectra collapse perfectly. This is due to the internal intermittency effects (\citet{Meyers2008}).       

\begin{figure}
\centering
	\subfigure{\label{fig:appriori:D111DNormalKolmoVSkxatDifferentTimesHMcase_incomp_corr.e}\includegraphics[width=0.3\textwidth]{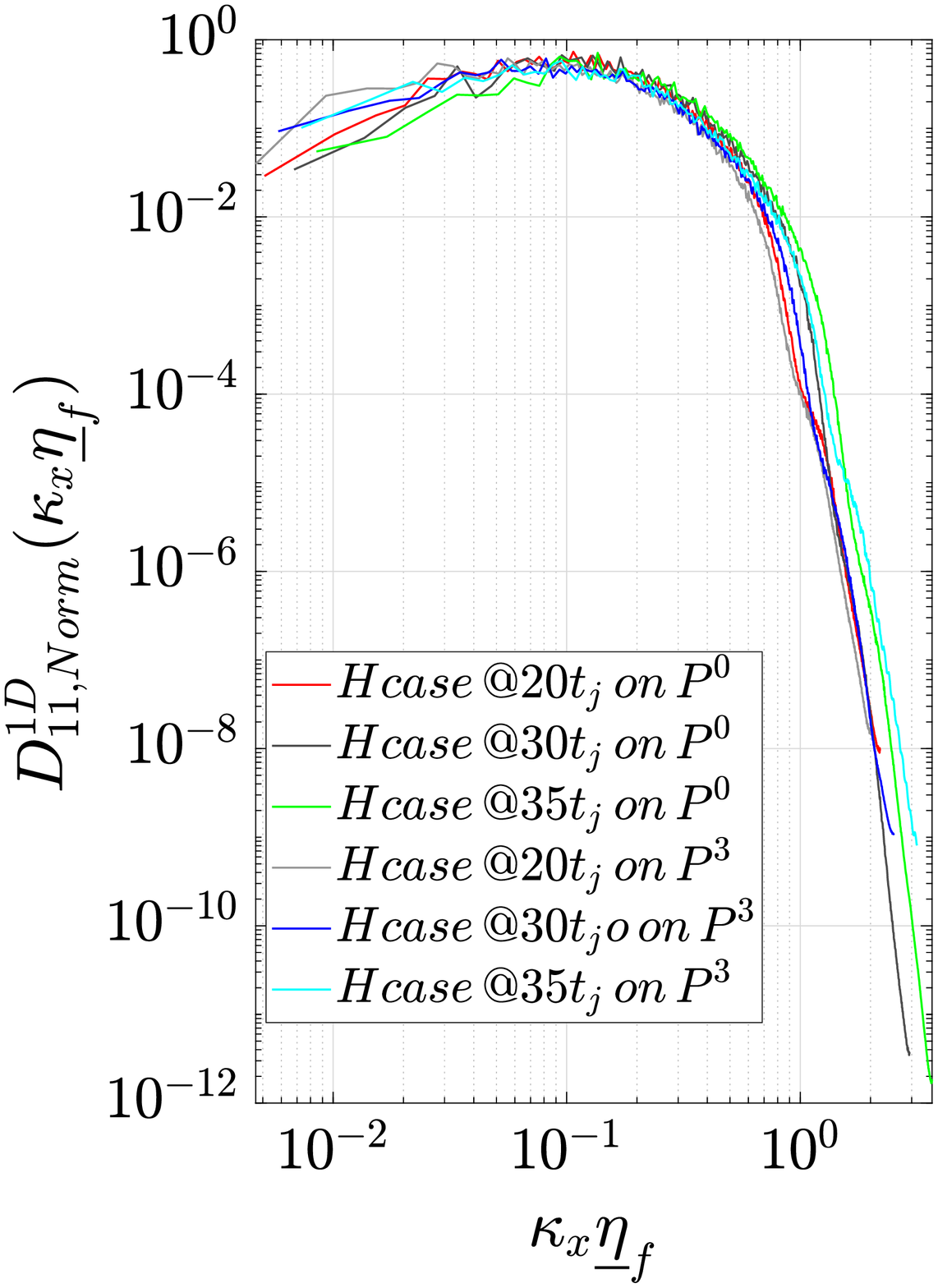}}
	\subfigure{\label{fig:appriori:D111DNormalKolmoVSkxatDifferentTimesHMcase_incomp_corr.f}\includegraphics[width=0.3\textwidth]{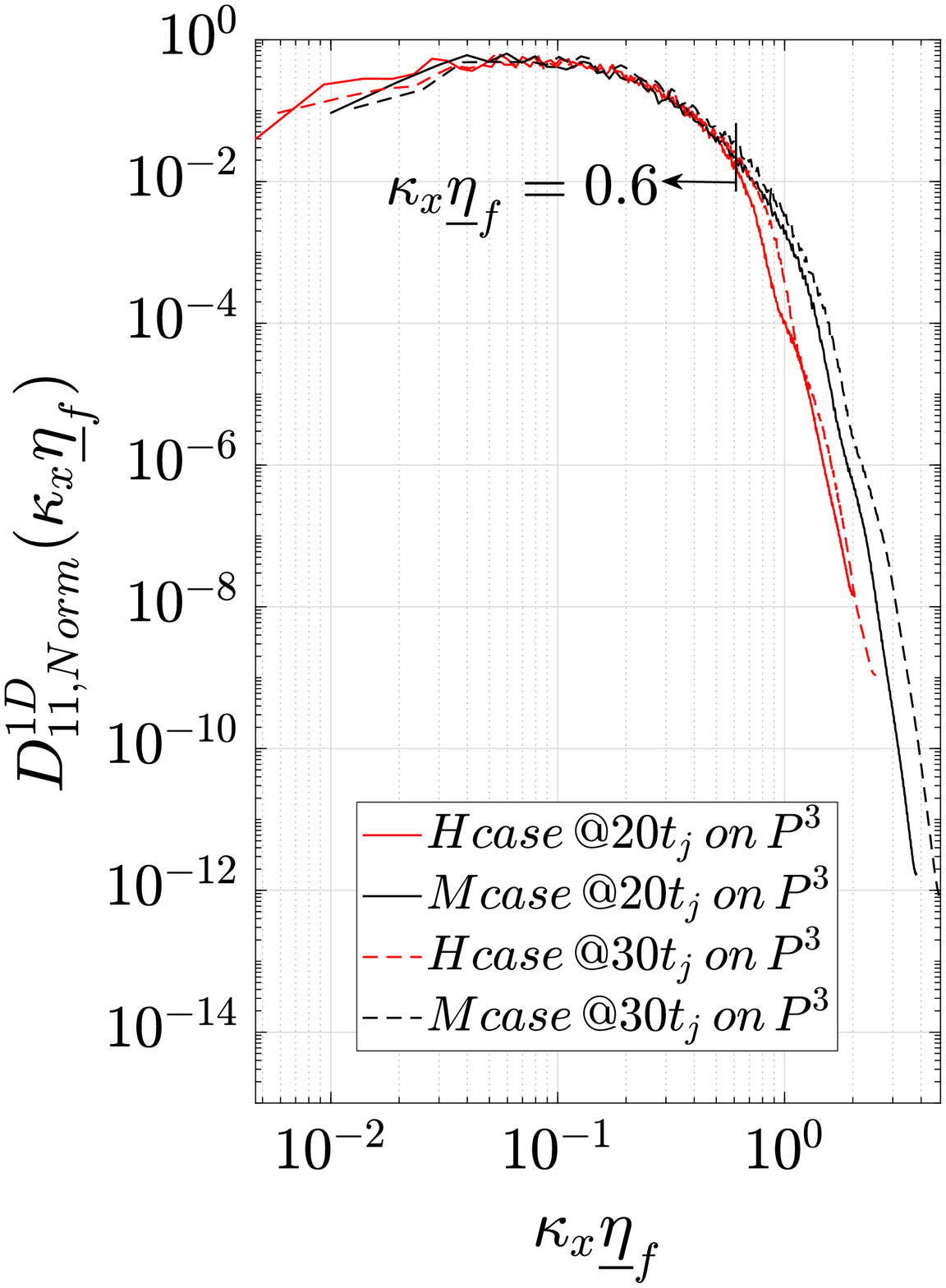}}		
	\caption{1D normalized longitudinal dissipation spectra, (\textbf{left}) on different planes and at different times for case H; (\textbf{right}) On $P^3$ for cases H and M at different times.}
	\label{fig:appriori:D111DNormalKolmoVSkxatDifferentTimesHMcase_incomp_corr}%
\end{figure}

The normalized dissipation spectra are also shown in figure~\ref{fig:appriori:D111DNormalKolmoVSkxatDifferentTimesHMcase_incomp_corr} in different conditions. Like before, for case H, up to $\kappa_x\underline{\eta}_f\approx 0.6$ the collapse is good, while for case M (the case with a higher resolution) the range extends to  $\kappa_x\underline{\eta}_f\approx 1.4$. The spectra extracted from both cases collapsed well up to the point where the change of slope of case H begins (see the right plot in figure~\ref{fig:appriori:D111DNormalKolmoVSkxatDifferentTimesHMcase_incomp_corr}). It is reasonable to expect that by increasing the resolution of case H, one could see an extended range of the collapse. Nevertheless, it should be considered that the amount of dissipation beyond $\kappa_x\underline{\eta}_f\approx 0.6$ is low (see the right plot in  figure~\ref{fig:appriori:D111DVSkxat0099Hcase}).

%\begin{figure*}
%\centering
%	\begin{subfigure}{0.3\textwidth}
%	\includegraphics[width=\textwidth]{D111DVSkxat0099Hcase_incomp_corr.eps}
%	\caption{\label{fig:appriori:D111DVSkxat0099Hcase.a}}
%	\end{subfigure}
%		\begin{subfigure}{0.3\textwidth}
%	\includegraphics[width=\textwidth]{D111DNormalKolmoVSkxat0099Hcase_incomp_corr.eps}
%	\caption{\label{fig:appriori:D111DVSkxat0099Hcase.b}}
%		\end{subfigure}
%	\caption{1D longitudinal dissipation spectra across the reacting shear layer at$t=20t_j$ for case H, (\textbf{a}) Non-normalized spectra; (\textbf{b}) Normalized using Favre averaged $\eta$ and $\varepsilon$.}
%	\label{fig:appriori:D111DVSkxat0099Hcase}%
%\end{figure*}	 

\section{Model Spectrum}\label{sec:appriori:modelSpect}
Pope's model (\cite{Pope2000turbulent}) for the 3D energy spectrum reads: 
\begin{equation}\label{eq:turb:Ek3DPope2}
E_{Model,Norm}(\kappa \underline{\eta}) = C_K  \left(\kappa \underline{\eta} \right)^{-5/3} f_\eta(\kappa\underline{\eta}) f_L(\kappa L),
\end{equation}
%where the integral range multiplier, $f_L(\kappa L)$ is active in low wavenumber ranges, viz. $f_L(\kappa L) = ( \dfrac{\kappa L}{( \left(\kappa L \right)^2 +c_L)^{1/2} } )^{5/3+P_0}$.  
The model has been proposed for constant density flows and includes the Kolmogorov spectrum for the inertial range, an integral range multiplier, $f_L(\kappa L)$, for low and a dissipation range multiplier, $f_\eta(\kappa\underline{\eta})$ for high wavenumbers. 
%Further it does not have effect on the shape of the model in these ranges. 
In the current work, the conventional form of $f_L(\kappa L)$, viz. $f_L(\kappa L) = ( \dfrac{\kappa L}{( \left(\kappa L \right)^2 +c_L)^{1/2} } )^{5/3+P_0}$, with same coefficients as non-reactive flows, i.e.,  $P_0=2$ and $c_L=6.78$ will be used since the wavenumber ranges of interest are the inertial and near dissipative ranges.
%($C_K  \left(\kappa \underline{\eta} \right)^{-5/3}$), 
%the $-5/3$ behaviour was already observed. Further, there are many evidences e.g., \citet{Kolla2014,Knaus2009} of existence of such a scaling in turbulent flames consistent with the non-reactive flows. 
Pope proposed $f_\eta(\kappa\underline{\eta})$ as:
\begin{equation}\label{eq:turb:feta2}
f_\eta(\kappa \underline{\eta}) = exp [-\beta \left( \left(\kappa \underline{\eta} \right)^4 +c_\eta^4  \right)^{1/4} - c_\eta ] ,
\end{equation}
with experimentally obtained constants of $\beta=5.2$ (\citet{saddoughi1994local}) and $ c_\eta =0.4$. As shown in the previous section we found a steeper exponential drop-off with  $\beta\approx7.2$. 
%for case H in range $0.1\leq\kappa_x \underline{\eta}_f\leq0.8$. This range is wider in case  M, i.e., $0.1\leq\kappa_x \underline{\eta}_f\leq1.8$ and $0.1\leq\kappa_x \underline{\eta}_f\leq1.4$, respectively (see Appendix~\ref{appendix:spectra}). So we use this value in the modified model. 
%As can be seen the exponential drop-off form in the dissipation range is more complicated than the simple form of Eq.~\ref{eq:appriori:KraichnanModel} already discussed, however, keeping the same exponential drop-off factor (if $c_\eta=0$). This must be used with care when comparing with the extracted spectra from DNS. 
%It is unclear how the 1D energy spectrum (i.e., $E_{1D}$) can be compared to 3D energy spectrum (i.e., $E$) as was done in \citet{Kolla2016} or \citet{Knaus2009}. 
The 3D model energy spectrum must be converted to a 1D spectrum to be compared to, e.g., $E_{11}$. If isotropy is assumed, one can derive the 1D velocity (e.g., streamwise) spectrum from the 3D energy spectrum (\citet{Monin1971statistical}):
\begin{equation}\label{eq:appriori:3DTo1Dspectrum}
E_{11,Model}^{1D}(\kappa_x)=\int_{\kappa_{x}}^{\infty} \left(1-\dfrac{\kappa_{x}^2}{\kappa^2}\right)\dfrac{E_{Model}(\kappa)}{\kappa}d\kappa
\end{equation}

\begin{figure}
	\centering			
	\includegraphics[width=0.5\textwidth]{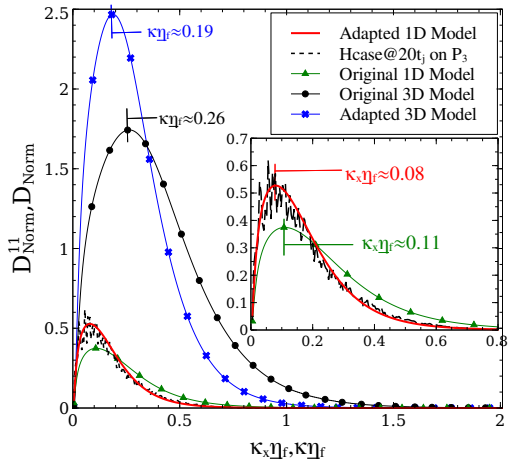}
	\caption{Different dissipation spectra on $\underline{Z}_f=Z_{st}$ plane at $t=20t_j$ for case H compared to model spectra. The inset is the zoom view of the left corner.}
	\label{fig:appriori:D11VSModel}%
\end{figure}

The dissipation spectrum reported in linear scales (figure~\ref{fig:appriori:D11VSModel}) shows interesting features. In this figure, five different normalized velocity dissipation spectra are plotted. The black line with circle markers is the 3D dissipation model of Pope with its peak at $\kappa \underline{\eta}_f\approx0.26$. It is clear that the 1D dissipation spectrum of stream-wise velocity fluctuations (the black dashed line) does not follow the model.  
%The model is for the {TKE} dissipation and the {DNS} data is the dissipation of the energy of velocity fluctuations in the stream-wise direction ($1/2u_1^{\prime\prime}u_1^{\prime\prime}$). 
%The true plot to be compared to the DNS data is 
The 1D form of the model computed by Eq.~\ref{eq:appriori:3DTo1Dspectrum} and plotted in  green with triangle markers is the one that should be compared with the DNS. 
The peak of the 1D model dissipation spectrum occurs at $\kappa_x \underline{\eta}_f\approx0.11$. This value is in agreement with the observations in experiments (e.g. \citet{saddoughi1994local}) and {DNS} (\citet{Ganapathisubramani2008}) of non-reactive flows. However, it is observed that the peak of the 1D dissipation spectrum of the non-premixed jet flames considered in this study is at a lower wavenumber ($\kappa_x\underline{\eta}_f\approx0.08$). Approximately the same value was observed in the DNS of premixed flames (see figure 9 in \citet{Kolla2016}). 
%It seems that in that paper it was compared with the peak location of the 3D model ($\kappa \underline{\eta}_f\approx0.26$) and was concluded that there is a large deviation from the results with respect to non-reactive model of Pope. 
%Now one may see that the correct value to be compared with is     $\underline{\eta}_f\approx0.11$. 
The red line is an adapted 1D model using the observed $C_K^{1D}\approx 2.3$, $\beta \approx7.2$ and fitted $c_\eta=0.28$. Finally, if the 3D model is plotted using these new coefficients, the result is the blue line with cross markers. 
%Now the blue and magenta lines must be compared. 
%The black line with circles is the 3D model spectrum for non-reactive flows and the one with cross markers has been obtained using the current non-premixed jets {DNS} databases. 
The peak of the new 3D model is at $\kappa \underline{\eta}_f\approx0.19$ instead of $\kappa \underline{\eta}_f\approx0.26$. Further, it is less extended to the far dissipation range ($\kappa \underline{\eta}_f>1$). Note that the areas under the two curves are equal.       
\section{Summary and Conclusions}
In this study the velocity and its dissipation spectra in non-premixed jet flames were analysed using two DNS databases of temporally evolving syngas jet flames experiencing high levels of local extinction followed by re-ignition. In each database, three different times were selected (6 in total) corresponding to the events of maximum local extinction and re-ignition. 
%
%The objective was to check Kolmogorov's scaling laws (with Favre averaged quantities) and to see if the spectra, specifically the dissipation spectra, follow the model spectrum of Pope (proposed for non-reactive flows). 
%
Consistent with \citet{Knaus2009} (non-premixed jet flames DNS) and \cite{Kolla2014} (premixed jet flames DNS), it was observed that using Favre averaged Kolmogorov's length scale and energy dissipation rate, namely $\underline{\eta}_f$ and $\underline{\varepsilon}_f$, the velocity spectra calculated on different planes across the reacting shear layers and in different flame dynamics, collapse very well on a curve in the inertial and near dissipation ranges before the discussed inflection occurs. 
%In the dissipation range the collapse can be improved using a cut-off scale defined by the inverse of a wavenumber corresponding to $2\%$ of the dissipation spectra peaks ($\lambda_{\beta}$). 
%To reach a perfect collapse of the spectra deep into the dissipation range $\kappa_x\underline{\eta}_f\gg1$, it was hypothesized that some information of chemistry may needed, however, in non-premixed flames it is not easy to define a characteristic length scale of the flame like what is done in premixed flames.
The $\kappa^{-5/3}$ power law was observed in the inertial range, with the constant of proportionality of $C_{K}=2.3$ instead of $1.5$. 
%The difference between the value of $C_{K}=2.3$ and the conventional $C_K=1.5-1.7$  \citet{Ishihara2009}, is probably due to the low $Re_{\lambda}$ and so the very limited inertial range in the current {DNS} databases \citet{Ishihara2009,ishihara2016energy}. 
In the dissipation range, the exponential drop-off factor of the spectra is steeper than the corresponding non-reactive flows. It was observed that  $\exp(\beta\kappa)$ scaling exists with $\beta=7.2$ instead of $\beta=5.2$. The 1D dissipation spectra do not follow the standard 1D model dissipation spectrum. Keeping the functional form of the model unchanged and using the observed values of $C_{K}=2.3$, $\beta=7.2$, and a fitting value of $c_\eta=0.28$, the agreement of the model spectrum and the dissipation spectra from DNS is highly improved. Future works will focus on extending the investigations on the drop-off factor, using reactive DNS databases with higher Re and higher resolutions. It is of interest to see if $\beta=7.2$ is a common behaviour in reactive jets or not. Moreover, using larger datasets, one may improve the fitting of $c_\eta$. The outcomes of this study can be applied in the modelling of turbulent reactive flows, either in turbulence modelling (e.g. for RANS/LES) or combustion modelling (e.g. EDC model of \cite{Ertesvag2000}). 

%Next, the spectra were compared with the model spectrum of Pope (proposed for non-reactive flows). It was found that the functional form proposed by \citet{Pope2000turbulent} for non-reactive flows agrees well with the velocity and the dissipation spectra of reactive flows studied in this work, if the values of $\mathrm{C_K}$ and $\beta$ mentioned above is used in the model. 
%We will use this adapted version of the model spectrum in Sec.~\ref{sec::TCOI:CD2} to derive an analytical relation between the two coefficients in the {EDC}.           

This work has received funding from the European Union's Horizon 2020 research and innovation programme under the Marie Sklodowska-Curie grant agreement No 643134. We would like to thank Professor Evatt Hawkes from The University of New South Wales for providing us the DNS databases and also Professor Dominique Th\'{e}venin from the University of Magdeburg and Dr. Christiane Zistl to provide us ANAFLAME codes.

\bibliographystyle{jfm}
\bibliography{Shamooni-Spectra-manuscript}

\end{document}